\input harvmac
\def\bigZ{Z\!\!\!Z}
\def\Wbar{{\bar W}}
\def\Phibar{\bar\Phi}
\lref\dkmone{N. Dorey, V.V. Khoze and M.P. Mattis, \it Multi-instanton
calculus in $N=2$ supersymmetric gauge theory\rm, 
Phys.~Rev.~D54 (1996) 2921, hep-th/9603136.}
\lref\Seiberg{ N. Seiberg, Phys. Lett. B206 (1988) 75. }
\lref\ADHM{ M.  Atiyah, V.  Drinfeld, N.  Hitchin and
Yu.~Manin, Phys. Lett. A65 (1978) 185. }
\lref\Corrigan{ E. Corrigan, P. Goddard and S. Templeton,
Nucl. Phys. B151 (1979) 93; \hfil\break
   E. Corrigan, D. Fairlie, P. Goddard and S. Templeton,
    Nucl. Phys. B140 (1978) 31.}
\lref\Osborn{H. Osborn, Ann. Phys. 135 (1981) 373. }
\lref\DineSeib{M. Dine and N. Seiberg, \it Comments
on Higher Derivative Operators in Some SUSY Field Theories\rm, hep-th/9705057.}
\lref\PW{A. Pickering and 
P. West, \it The One Loop Effective Superpotential and 
Nonholomorphicity\rm, Phys. Lett. B383 (1996) 54, hep-th/9604147.}
\lref\PP{J. Polchinski and P. Pouliot, \it Membrane Scattering
with M-momentum Transfer\rm, hep-th/9704029.}
\lref\dkmeight{N. Dorey, V.V. Khoze and M.P. Mattis, \it Multi-Instantons,
Three-Dimensional Gauge Theory and the Gauss-Bonnet-Chern 
Theorem\rm,  hep-th/9704197.}
\lref\BFSS{T. Banks, W. Fischler, N. Seiberg and L. Susskind, \it Instantons,
Scale Invariance and Lorentz Invariance in Matrix Theory\rm, hep-th/9705190.}
\lref\LGRU{U. Lindstrom, F. Gonzalez-Rey, M. Rocek and R. von Unge,
\it On $N=2$ Low-Energy Effective Actions\rm, 
Phys. Lett. B388 (1996) 581, hep-th/9607089.}
\lref\Yung{A. Yung, \it Instanton-induced Effective Lagrangian in
the Seiberg-Witten Model\rm, Nucl. Phys. B485 (1997) 38, 
hep-th/9605096; and 
\it Higher Derivative Terms in the Effective Action of $N=2$ SUSY QCD
from Instantons\rm, hep-th/9705181.}
\lref\dkmfour{N. Dorey, V.V. Khoze and M.P. Mattis, \it Multi-instanton
calculus in $N=2$ supersymmetric gauge theory.
II. Coupling to matter\rm, Phys.~Rev.~D54 (1996) 7832, hep-th/9607202.}
\lref\dkmfive{N. Dorey, V.V. Khoze and M.P. Mattis, \it On $N=2$
supersymmetric QCD with $4$ flavors\rm,  hep-th/9611016.}
\lref\dkmsix{N. Dorey, V.V. Khoze and M.P. Mattis, \it On mass-deformed $N=4$
supersymmetric Yang-Mills theory\rm,  
Phys. Lett. B396 (1997) 141, hep-th/9612231.}
\lref\Henningson{M. Henningson, 
\it Extended Superspace, Higher Derivatives and 
$SL(2,\bigZ)$ Duality\rm, Nucl. Phys. B458 (1996) 445, hep-th/9507135.}
\lref\dWGR{B. de Wit, M.T. Grisaru and M. Rocek, \it Nonholomorphic
Corrections to the One-loop $N=2$ Super Yang-Mills Action\rm, 
Phys. Lett. B374 (1996) 297, hep-th/9601115.}
\lref\Gates{G. Sierra and P.K. Townsend, in \it Supersymmetry and Supergravity
1983\rm, proc. XIXth Winter School, Karpacz, ed. 
B. Milewski, World Scientific 1983, p.396;
\hfil\break
S.J. Gates, Nucl. Phys. B238 (1984) 349;\hfil\break
  B. de Wit, P.G. Lauwers, R. Philippe, S.-Q. Su and A. Van Proeyen, 
  Phys. Lett. B134 (1984) 37.}
\lref\GSW{R. Grimm, M. Sohnius and J. Wess, Nucl. Phys. B133 (1978) 275.}
\lref\Wil{M.A. Shifman and A.I. Vainshtein, Nucl. Phys. B277 (1986) 456;
          \hfil\break
          M. Dine and Shirman, Phys. Rev. D50 (1994) 5389, hep-th/9405155.}
\lref\SW{N. Seiberg and E. Witten, 
{\it Electric-magnetic duality, monopole
condensation, and confinement in $N=2$ supersymmetric Yang-Mills theory}, 
Nucl. Phys. B426 (1994) 19, (E) B430 (1994) 485  hep-th/9407087; and
{\it Monopoles, duality and chiral symmetry breaking
in $N=2$ supersymmetric QCD}, 
Nucl. Phys B431 (1994) 484,  hep-th/9408099.}
\def\susic{supersymmetric}
\def\sigmabar{\bar\sigma}
\def\quarter{{\textstyle{1\over4}}}

\def\fourth{\quarter}
\def\sqrtwo{\sqrt{2}}
\def\wbar{\bar w}
\def\Ubar{\bar U}
\def\bbar{\bar b}
\def\susy{supersymmetry}
\def\v{{\rm v}}
\def\vbar{\bar{\rm v}}
\def\F{{\cal F}}
\def\H{{\cal H}}
\def\V{{\cal V}}
\def\B{{\cal B}}
\def\Vbar{\bar{\cal V}}
\def\M{{\cal M}}
\def\N{{\cal N}}
\def\R{{\cal R}}
\def\K{{\cal K}}
\def\Rtilde{\tilde{\cal R}}
\def\Ktilde{\tilde{\cal K}}
\def\G{{\cal O}}
\def\Gbar{\bar{\cal O}}
\font\authorfont=cmcsc10 \ifx\answ\bigans\else scaled\magstep1\fi
\divide\baselineskip by 10
\multiply\baselineskip by 9
{\divide\baselineskip by 4
\multiply\baselineskip by 4
\def\prenomat{\hbox{hep-th/9706007}}
\Title{$\prenomat$}{\vbox{\centerline{Instantons, Higher-Derivative Terms,}
\vskip .05in
\centerline{and Nonrenormalization Theorems}
\vskip .07in
\centerline{in Supersymmetric Gauge Theories}}}
\centerline{\authorfont N. Dorey$^1$, V.V. Khoze$^2$, M.P. Mattis$^3$, 
M.J. Slater$^2$, and W.A. Weir$^2$}
\bigskip
\bigskip
\centerline{\sl $^1$Physics Department, University of Wales Swansea}
\centerline{\sl Swansea SA2$\,$8PP UK}
\bigskip
\centerline{\sl $^2$Centre for Particle Theory,
University of Durham}
\centerline{\sl Durham DH1$\,$3LE UK}
\bigskip
\centerline{\sl $^3$Theoretical Division T-8, Los Alamos National Laboratory}
\centerline{\sl Los Alamos, NM 87545 USA}
\vskip .3in

\def\quarter{{\textstyle{1\over4}}}
\noindent
We discuss the contribution of
 ADHM multi-instantons to the higher-derivative
terms in the gradient expansion along the Coulomb branch of $N=2$ and
$N=4$ supersymmetric $SU(2)$ gauge theories. In particular, using simple
scaling arguments, we confirm the Dine-Seiberg nonperturbative
nonrenormalization theorems  for the 4-derivative/8-fermion term in
the two finite theories ($N=4$, and $N=2$ with $N_F=4$).
\vskip .1in
\Date{\bf June 1997} 
\vfil\break
}
\bf 1. Introduction. \rm
Thanks largely to the work of Seiberg and Witten \refs{\Seiberg,\SW},
much is now understood about spontaneously broken 4-dimensional
$SU(2)$ gauge theories
with extended \susy. One aspect that has
received  extensive study is the structure of the (suitably defined
\Wil) Wilsonian effective action along the  Coulomb
branch in which the models retain an unbroken $U(1)$ gauge symmetry.
For energies well below the symmetry
breaking scale, the dynamics of the massless $U(1)$ modes
 may be analyzed in a (supersymmetrized) gradient
expansion for the scalar fields, 
the form of which is constrained by both gauge invariance and
$N=2$ \susy. In particular the leading 2-derivative/\hbox{4-fermion} term
is expressed in terms of a holomorphic
 object $\F(\Psi)$ known as the prepotential \refs{\Gates,\Seiberg}:
\def\Ltwoderiv{{\cal L}_{2\hbox{-}\rm deriv}}
\def\Lfourderiv{{\cal L}_{4\hbox{-}\rm deriv}}
\def\Psibar{\bar\Psi}
\eqn\Ltwoderivdef{\Ltwoderiv\ =\ {1\over4\pi}\,
{\rm Im}\int d^4\theta\,\F(\Psi)\ ,}
where $\Psi$ denotes the massless $N=2$  abelian chiral superfield \GSW.
Equation \Ltwoderivdef\
 is an $N=2$ $F$-term as it involves integration over half the
$N=2$ superspace. 
In contrast, the next-leading term, involving
four derivatives and up to eight fermions, is an $N=2$ $D$-term
\refs{\Henningson,\dWGR}:
\eqn\Lfourderivdef{\Lfourderiv\ =\ \int d^4\theta d^4\bar\theta\,
\H(\Psi,\Psibar)}
where $\H$ is a real function of its arguments. This
\susic\ gradient expansion has been systematized by Henningson
\Henningson. 

By exploiting holomorphicity, together with electric-magnetic
duality, Seiberg and Witten were able to produce the exact quantum
solution for $\F$ in a variety of
$SU(2)$ models \SW. (There are however some interpretational caveats in
the case of the two finite models, namely the $N=2$ model with $N_F=4$
flavors of quark hypermultiplets \dkmfive, and the $N=4$ model
\dkmsix.) 
In contrast, comparatively little is known in general about the
 function $\H$ \refs{\Henningson-\dWGR,\PW\LGRU-\Yung}, since it is
real rather than holomorphic (although it does respect duality \Henningson). 
But for the two finite models, it turns out that exact statements can
nevertheless be made. In particular, Dine and Seiberg have recently
argued that in both these  cases
 $\Lfourderiv$ is one-loop exact: the one-loop result
receives corrections neither from higher orders in perturbation
theory, nor from nonperturbative physics such as instantons \DineSeib.

In this note, we discuss the contribution of Atiyah-Drinfeld-Hitchin-Manin
(ADHM) multi-instantons \refs{\ADHM\Corrigan\Osborn-\dkmone}
to these higher terms in the gradient expansion. Our principal
result (Eq.~(25) below) is a formal expression for the leading
semiclassical contribution
of the pure \hbox{$n$-instanton} (or pure $n$-antiinstanton) sector to $\H$,
expressed as a finite-dimensional integral over the bosonic and fermionic
collective coordinates of the supersymmetrized ADHM multi-instanton.
As a simple illustration, we calculate the 1-instanton contribution
to $\H$ in the case of pure $N=2$ SYM theory, and reproduce an earlier
result of Yung's \Yung. 
When $n>2$ the expression (25) is truly a  formal one only, since the measure
for this integration is not currently known \refs{\Osborn,\dkmone}.
Nevertheless, for the finite model with $N=2$ and $N_F=4$,
we can verify, using a simple scaling argument, the
vanishing of these multi-instanton contributions 
to $\H$, for all values of the topological charge $n$. 
A slightly modified scaling argument extends this null result to the
$N=4$ model as well. Thus the Dine-Seiberg nonperturbative
nonrenormalization theorems are built into the ADHM instanton calculus.

\bf 2. Instanton basics. \rm
The massless $U(1)$ 
modes of $N=2$ \susic\ $SU(2)$ gauge theory along the Coulomb
branch consist of a photon field $v_m,$ two Weyl spinors $\lambda$ and
$\psi,$ and a complex scalar $A\,$; these assemble into a single neutral
massless $N=2$ chiral superfield $\Psi$. 
We start by reviewing the instanton representation of the prepotential
$\F\,$ \dkmfour; a straightforward extension of these methods
 will then yield an analogous
formula for $\H$. Let us expand $\Ltwoderiv$ in component fields, and focus
(as in Sec.~5 of \dkmone) on the following three effective vertices:
\eqn\vertices{\Ltwoderiv\ \ \supset\ \ {1\over4\pi}\,
\sum_{k=1,2,3}\V^k\circ\F(\v)
\ +\ \rm H.c.\ ,}
where $\v$ denotes the VEV of the Higgs field $A$, and 
\def\sst{\scriptscriptstyle}
\def\vsd{v^{\sst\rm SD}}
\def\vasd{v^{\sst\rm ASD}}
\def\psibar{\bar\psi}
\def\lambdabar{\bar\lambda}
\def\dalpha{{\dot\alpha}}
\def\dbeta{{\dot\beta}}
\def\dgamma{{\dot\gamma}}
\def\ddelta{{\dot\delta}}
\eqna\Vdef
$$\eqalignno{
\V^1\ &=\ {i\over4}\,\big(\vsd_{mn}\big)^2{\partial^2\over\partial\v^2}
\ ,&\Vdef a
\cr
\V^2\ &=\ {i\over2\sqrtwo}\,\psi\sigma^{mn}\lambda\vsd_{mn}
{\partial^3\over\partial\v^3}
\ ,&\Vdef b
\cr
\V^3\ &=\ -{i\over8}\,\psi^2\lambda^2
{\partial^4\over\partial\v^4} \ .
&\Vdef c
}
$$
The superscript $\scriptstyle\rm SD$ indicates the self-dual part of the
gauge field strength $v_{mn}.$\foot{In 
Minkowski space the self-dual and anti-self-dual
components of $v_{mn}$ are projected out using
$\vsd_{mn}\ =\ \quarter\big(\eta_{mk}\eta_{nl}-\eta_{ml}
\eta_{nk}+i\epsilon_{mnkl}\big)v^{kl}$ and $\vasd_{mn}=(\vsd_{mn})^*$,
where $\epsilon_{0123}=-\epsilon^{0123}=-1$. Also, since
$\sigma^{mn}=\fourth\,\sigma^{[\,m}\sigmabar^{n\,]}$ and
$\sigmabar^{mn}=\fourth\,\sigmabar^{[\,m}\sigma^{n\,]}$ are
self-dual and anti-self-dual, respectively, it follows that
$\sigma^{mn\ \beta}_{\phantom{mn}\alpha}v_{mn}
=\sigma^{mn\ \beta}_{\phantom{mn}\alpha}\vsd_{mn}$
and
$\sigmabar^{mn\dot\alpha}_{\phantom{mn\alpha}\dot\beta}v_{mn}
=\sigmabar^{mn\dot\alpha}_{\phantom{mn\alpha}\dot\beta}\vasd_{mn}$.
Here  $\sigma_m$ and $\bar\sigma_m$ are spin matrices
in Wess and Bagger conventions; see \dkmone\ for a complete set of our
SUSY and ADHM conventions.}
We will call such vertices ``holomorphic'' as the fields
$\psi,$ $\lambda$ and $\vsd_{mn}$ live in the chiral
superfield $\Psi$ rather than in $\Psibar.$
To extract the (multi-)instanton contribution to these three holomorphic
vertices, one analyzes, respectively, the three \it anti\rm holomorphic
Green's functions
$\langle\Gbar^k(x_1,\ldots,x_{k+1})\rangle,$ $k=1,2,3,$ with
\eqna\Gbardef
$$\eqalignno{
\Gbar^1(x_1,x_2)\ &=\ \vasd_{mn}(x_1)\,\vasd_{pq}(x_2)
\ ,&\Gbardef a
\cr\Gbar^2(x_1,x_2,x_3)\ &=\ \psibar_\dalpha(x_1)\,\vasd_{mn}(x_2)\,
\lambdabar_\dbeta(x_3)\ ,&\Gbardef b\cr
\Gbar^3(x_1,x_2,x_3,x_4)\ &=\ \psibar_\dalpha(x_1)\,\psibar_\dbeta(x_2)
\,\lambdabar_\dgamma(x_3)\,\lambdabar_\ddelta(x_4)\ .&\Gbardef c}
$$
In the semiclassical approximation these field insertions are simply replaced
by their  values in the classical (multi-)instanton background, 
projected onto the unbroken $U(1)$ direction in color space, then
integrated over all bosonic and fermionic instanton collective coordinates,
which we now briefly review.

In the ADHM formulation, the general multi-instanton solution of topological
charge $n$ in $SU(2)$ gauge theory may be parametrized by an $(n+1)\times n$
quaternion-valued\foot{We use quaternionic notation, e.g.,
$x=x_{\alpha\dalpha}=x_m\sigma^m_{\alpha\dalpha}$ and
$\bar x=\bar x^{\dalpha\alpha}=x^m\bar\sigma_m^{\dalpha\alpha}.$}
collective coordinate matrix $a_{\alpha\dalpha}$, while the adjoint
fermionic zero modes for the gaugino $\lambda^\gamma$ and Higgsino
$\psi^\gamma$ are expressed in terms of $(n+1)\times n$ Weyl-valued
collective coordinate matrices $\M^\gamma$ and $\N^\gamma$, respectively
\refs{\ADHM-\dkmone}:
\eqn\bcanonical{
a_{\alpha\dalpha} =
\pmatrix{w_{1\alpha\dalpha}&\cdots&w_{n\alpha\dalpha}
\cr{}&{}&{}\cr
{}&a'_{\alpha\dalpha}&{}\cr{}&{}&{}}\ ,\ \
\M^\gamma = \pmatrix{\mu_1^\gamma&\cdots&\mu_n^\gamma
\cr{}&{}&{}\cr
{}&\M^{\prime\gamma}&{}\cr{}&{}&{}}
\ ,\ \
\N^\gamma = \pmatrix{\nu_1^\gamma&\cdots&\nu_n^\gamma
\cr{}&{}&{}\cr
{}&\N^{\prime\gamma}&{}\cr{}&{}&{}} }
with 
$a_{\alpha\dalpha}'=a_{\alpha\dalpha}^{\prime T},$
$\M_\gamma'=\M_\gamma^{\prime T},$ and
$\N_\gamma'=\N_\gamma^{\prime T}.$ Furthermore the matrices $a,$ $\M$
and $\N$ are subject to a set of algebraic constraints which may be used,
for example, to eliminate the off-diagonal elements of the $n\times n$
submatrices $a',$ $\M'$ and $\N'.$ This leaves $4\times 2n$ independent
scalar degrees of freedom in $a$ (i.e., the $w_k$ and the diagonal
elements of $a'$), and likewise $2\times 2n$ independent Grassmann
degrees of freedom in each of $\M$ and $\N.$ Of these, the `trace'
components of $a',$ $\M'$, and $\N'$, respectively,
 play a special role: that of the
position $(x_{0\alpha\dalpha}\,,\,\xi^1_\alpha\,,\,\xi^2_\alpha)$ of the
multi-instanton in $N=2$ superspace (see Eq.~(8.1) of Ref. \dkmone).

The above describes the collective coordinate space for pure $N=2$ $SU(2)$
gauge theory. When $N_F$ (massive) flavors of $N=2$ ``quark hypermultiplets''
in the fundamental representation of the gauge group are coupled in,
one needs $2nN_F$ additional Grassmann collective coordinates for the
fundamental fermion zero modes \Corrigan; following \dkmfour, we label these
$\K_{ki}$ and $\Ktilde_{ki}$ with $k=1,\cdots,n$ and $i=1,\cdots,N_F$.
Alternatively, if a single (massive) adjoint hypermultiplet is coupled in,
there are new adjoint fermion zero modes requiring $8n$ 
Grassmann degrees of freedom; following \dkmsix, these may be taken to live
in the $(n+1)\times n$ Weyl-valued collective coordinate matrices 
\def\rhotilde{\tilde\rho}
\eqn\newcanon{\R^\gamma = \pmatrix{\rho_1^\gamma&\cdots&\rho_n^\gamma
\cr{}&{}&{}\cr
{}&\R^{\prime\gamma}&{}\cr{}&{}&{}}
\ ,\ \
\Rtilde^\gamma = \pmatrix{\rhotilde_1^\gamma&\cdots&\rhotilde_n^\gamma
\cr{}&{}&{}\cr
{}&\Rtilde^{\prime\gamma}&{}\cr{}&{}&{}} \ ,}
which are subject to the same algebraic constraints as $\M$ and $\N.$

\def\Sinstn{S^n_{\rm inst}}
In previous work we have derived explicit formulae for the instanton
action $\Sinstn$ for arbitrary topological charge $n$, as functions
of these collective coordinates, as well as of the VEVs $\v$ and $\vbar$
and of the hypermultiplet masses $m_i\,$:
\eqn\Sinstdef{\Sinstn\ =\ \Sinstn\big(\,
\{a\}\,,\,\{\M\,,\,
\N\}\,;\,
\{\K\,,\,\Ktilde\}\hbox{ or }
\{\R\,,\,\Rtilde\}\,;\,
\v,\vbar\,;\,\{m_i\}\,\big)}
Here we will not actually
need these formulae\foot{See 
Eq.~(7.32) of \dkmone\ for the explicit expression in the case of pure $N=2$
SYM theory, Eq.~(5.20) of \dkmfour\ for the incorporation of $N_F$ (massive)
fundamental hypermultiplets, and Eq.~(15) of \dkmsix\ for the incorporation
of a single (massive) adjoint hypermultiplet.}; 
it will suffice to note some general features of $\Sinstn.$ 
To begin with (save for one special case discussed below, that of exact
$N=4$ SYM theory), $\Sinstn$
explicitly depends on all the Grassmann collective coordinates in the
problem \it except \rm
for the four exact $N=2$ SUSY modes $\xi^1_\alpha$ and $\xi^2_\alpha$,
$\alpha=1,2,$ described above. Recall the rules of Grassmann integration:
$\int d^2\xi^i\,(\xi^i)^2=1$ while $\int d^2\xi^i=0$. 
Thus, in order to saturate the $d^2\xi^1\,d^2\xi^2$ integration, 
one requires the explicit insertion of $\xi^i$-dependent 
component fields, e.g., the $\Gbar^k$ of Eq.~\Gbardef{}. The remaining
Grassmann integrations are saturated by pulling down the appropriate
power of $\Sinstn$ from the exponent. 

We illustrate these comments by focusing, first, on the $n$-instanton
contribution to the 4-antifermion Green's function \Gbardef c:
\def\Dmu{d\tilde{\mu}}
\def\Dppmu{D\hat{\mu}}
\def\LD{{\sst\rm LD}}
\def\ld{{\sst\rm LD}}
\eqn\fourfold{\eqalign{&\langle\psibar_\dalpha(x_1)\,\psibar_\dbeta(x_2)
\,\lambdabar_\dgamma(x_3)\,\lambdabar_\ddelta(x_4)\rangle
\,{\Big|}_{n\hbox{-}\rm
inst}\ \cong\ \cr&\int d^4x_0\,d^2\xi^1\,d^2\xi^2\,\int\Dmu\,
\psibar^\LD_\dalpha(x_1)\,\psibar^\LD_\dbeta(x_2)\,\lambdabar^\LD_\dgamma(x_3)
\,\lambdabar^\LD_\ddelta(x_4)\,\exp(-\Sinstn)\ .}}
Here $\Dmu$ stands for the properly normalized integration measure for all the
collective coordinates in the problem (bosonic and fermionic, adjoint and
fundamental) \refs{\Osborn-\dkmfour}, 
excepting the $N=2$ superspace position variables
$(x_{0\alpha\dalpha}\,,\,\xi^1_\alpha\,,\,\xi^2_\alpha)$ 
which have been written out explicitly.
As indicated, at leading order, $\psibar$ and $\lambdabar$ are
approximated by quantities $\psibar^\LD$ and $\lambdabar^\LD$ defined
as follows \refs{\Seiberg,\dkmone,\dkmfour}: 
first, one solves the Euler-Lagrange equations for
$\psibar$ and $\lambdabar$ in the classical background of the
ADHM multi-instanton with all its fermionic zero modes turned on (and
parametrized by the collective coordinates described above); next,
one projects the resulting $SU(2)$-valued configurations onto the unbroken
$U(1)$ direction in the color space (this is the direction parallel
to the adjoint VEV); and finally, one assumes that the insertion points
$x_i$ are far away from the instanton position $x_0$ and performs
a long-distance (LD) expansion. For all the $N=2$ models, 
the result of this 3-step procedure may be
expressed compactly as follows \dkmfour:\foot{
The three effective vertices \vertices\ in $\Ltwoderiv$ are precisely those for
which the tail of the instanton dominates the $d^4x_0$ integration; 
likewise for the nine effective
vertices (18) in $\Lfourderiv$.}
\eqna\LDfermions
$$\eqalignno{
\psibar^\LD_\dalpha(x_i)
\ &=\ i\sqrtwo\,\xi^{1\alpha}\,S_{\alpha\dalpha}
(x_i,x_0)\,{\partial\over\partial\v}\ +\ \cdots
&\LDfermions a
\cr
\lambdabar_\dalpha^\LD(x_i)
\ &=\ -i\sqrtwo\,\xi^{2\alpha}\,S_{\alpha\dalpha}
(x_i,x_0)\,{\partial\over\partial\v}\ +\ \cdots
&\LDfermions b
}$$
Here $S_{\alpha\dalpha}$ is the Weyl spinor propagator,
\eqn\Sprop{S_{\alpha\dalpha}(x_i,x_0)\ =\ \sigma^m_{\alpha\dalpha}
\partial_m\,G_0(x_i,x_0)\ ,\quad
G_0(x_i,x_0)\ =\ {1\over4\pi^2(x_i-x_0)^2}\ ,}
and the derivative $\partial/\partial\v$ acts on $\exp(-\Sinstn),$
with the understanding that
$\v$ and $\vbar$ are always to be treated as independent
variables.
The omitted terms in \LDfermions{} represent terms that fall off faster
than $(x_i-x_0)^{-3}$, as well as terms that are independent of
$\xi^1$ or $\xi^2$ and hence cannot saturate these integrations.
Note that in the models with hypermultiplets, $\psibar^\LD$ and
$\lambdabar^\LD$ as given in \LDfermions{} contain both linear and
trilinear terms in Grassmann variables (hence would be tricky to derive
using Feynman graphs rather than the methods of \dkmfour).
Substituting Eq.~\LDfermions{} into Eq.~\fourfold\
and performing the $\xi^i$ integrals yields
\eqn\newfourfold{\int d^4x_0\,
S^\alpha_{\ \dalpha}(x_1,x_0)S_{\alpha\dbeta}(x_2,x_0)
S^\gamma_{\ \dgamma}(x_3,x_0)S_{\gamma\ddelta}(x_4,x_0)
\,{\partial^4\over\partial\v^4}\int\Dmu\,e^{-\Sinstn}\ .}
This one recognizes as the position-space Feynman graph for the effective
4-fermion vertex $-{i\over32\pi}\,\psi^2\lambda^2\F_n''''(\v)$ with
\dkmfour:
\eqn\Fndef{\F_n(\v)\ \equiv\ \F(\v){\Big|}_{n\hbox{-}\rm
inst}\ =\ 8\pi i\int\Dmu\,e^{-\Sinstn}\ .}

Similarly, in order to generate the $n$-instanton contribution to the effective
vertices \Vdef{a\hbox{-}b} 
one analyzes the Green's functions \Gbardef{a\hbox{-}b}, respectively. These
require the long-distance expression for the anti-self-dual part of the field
strength \refs{\dkmfour}:\foot{The 
fact that the gauge field strength develops an anti-self-dual
piece in perturbation theory around the instanton is detailed in
Sec.~4.4 of \dkmone.}
\eqn\vmnld{v_{mn}^{\sst\rm ASD,LD}(x_i)\ =\ \sqrtwo\,\xi^1\sigma^{pq}\xi^2\,
G_{mn,pq}(x_i,x_0)\,{\partial\over\partial\v}\ +\ \cdots}
where $G_{mn,pq}$ is the gauge-invariant propagator of $U(1)$ field
strengths:
\eqn\Gmnpqdef{G_{mn,pq}(x_i,x_0)\ =\ \big(\,\eta_{mp}\partial_n\partial_q
-\eta_{mq}\partial_n\partial_p-\eta_{np}\partial_m\partial_q+
\eta_{nq}\partial_m\partial_p\,\big)G_0(x_i,x_0)\ .}
The omitted terms in \vmnld\ include terms that fall off faster than
$(x_i-x_0)^{-4}$, as well as terms containing fewer than two of the $\xi^i$
modes  and hence cannot saturate these integrations.
An important property of $G_{mn,pq}$ is that it only connects
$\vsd_{mn}$ to $\vasd_{pq}$ and vice versa (just as $S_{\alpha\dalpha}$
only connects $\lambda$ to $\lambdabar$, and $\psi$ to $\psibar$).
This property follows from the identity
\eqn\Gpropertya{\sigmabar^{pq\dalpha}{}_\dbeta\,
G_{mn,pq}(x)\ =\ {2\over\pi^2x^6}\,\bar x^{\dalpha\alpha}\,
\sigma_{mn\,\alpha}{}^\beta\,x_{\beta\dbeta}}
which implies
\eqn\Gpropertyb{0\ =\ 
\sigmabar^{mn}_{\dgamma\ddelta}\,\sigmabar^{pq}_{\dalpha\dbeta}\,
G_{mn,pq}(x)\ =\ 
\sigma^{mn}_{\gamma\delta}\,\sigma^{pq}_{\alpha\beta}\,G_{mn,pq}(x)\ .}
Now the Green's functions \Gbardef{a\hbox{-}b} may be calculated
as before, by substituting the long-distance expressions \vmnld\ and
\LDfermions{} into the collective coordinate integration, and performing
the $\xi^i$ integrals explicitly. Thanks to Eq.~\Gpropertyb, one indeed
recovers the effective vertices \Vdef{a\hbox{-}b}, with the $n$-instanton
contribution to the prepotential still given by Eq.~\Fndef\ as the
reader can check \dkmfour.

\bf 3. Multi-instanton contribution to $\H(\Psi,\bar\Psi).$ \rm
A straightforward extension of these methods gives useful information
about $\Lfourderiv$ too (as well as higher terms in the gradient
expansion). As before, it is useful to expand $\Lfourderiv$ in component
fields, and to focus on the following nine effective vertices\foot{In 
$N=1$ language these nine vertices are all contained in 
the last term in Eq.~(4.7) in \Henningson, which in our $d^2\theta$
conventions reads
$\fourth \int d^2\theta d^2\bar\theta\,W^2\Wbar^2
\partial^4 \H(\Phi,\Phibar) / \partial^2 \Phi \partial^2 \Phibar\,.$}:
\eqn\newvertices{\Lfourderiv\ 
\ \supset\ \ 4 \sum_{k,k'=1,2,3}\V^k\circ\Vbar^{k'}\circ \H(\v,\vbar) .}
Here $\H$ is the kernel in Eq.~\Lfourderivdef, the $\V^k$ are the
holomorphic vertices \Vdef{}, and the $\Vbar^k$ are their
Hermitian conjugates (e.g., $\Vbar^2=-{i\over2\sqrtwo}\,
\psibar\sigmabar^{mn}\lambdabar\,
\vasd_{mn}\,\partial^3/\partial\vbar^3\ $). Again as before, these nine
vertices are probed, respectively, by the nine
antiholomorphic$\,\times\,$holomorphic Green's functions
\def\bG{{\bf G}}
\eqn\newGreens{\bG^{k,k'}(x_1,\ldots,x_{k+1},
y_1,\ldots,y_{k'+1})\ =\
\langle
\Gbar^k(x_1,\ldots,x_{k+1})\,
\G^{k'}(y_1,\ldots,y_{k'+1})\rangle\ ,}
where the $\Gbar^k$ are given in \Gbardef{} and the $\G^{k'}$ are their
Hermitian conjugates, $k,k'=1,2,3$. 
We now need, in addition to Eqs.~\LDfermions{}
and \vmnld, the long-distance expansions of the fields $\psi_\alpha,$
$\lambda_\alpha$ and $\vsd_{mn}.$ These are easily derived from the full
$SU(2)$ expressions \refs{\ADHM,\Corrigan,\dkmone}:
\eqna\oldies
$$\eqalignno{(\vsd_{mn})^\dalpha{}_\dbeta\ &=\ 
4\Ubar^{\dalpha\alpha}\,b\,
\sigma_{mn\,\alpha}{}^\beta
\,f\,\bbar\,
U_{\beta\dbeta}
&\oldies a\cr
(\lambda_\alpha)^\dalpha{}_\dbeta\ &=\ 
\Ubar^{\dalpha\gamma}\M_\gamma f\,\bbar\, U_{\alpha\dbeta}\ -\
\Ubar^\dalpha{}_\alpha \,bf\M^{\gamma T}U_{\gamma\dbeta}
&\oldies b}$$
Here $\dalpha$ and $\dbeta$ are color $SU(2)$ indices, the ADHM quantities
$U,$ $f$ and $b$ are as defined in Sec.~6 of \dkmone, and $\M_\gamma$ is the
Grassmann collective coordinate matrix \bcanonical; for
$(\psi_\alpha)^\dalpha{}_\dbeta$, substitute $\N_\gamma$ for $\M_\gamma.$
Projecting onto the unbroken $U(1)$ direction (which we assume for 
definiteness to lie in
the $\tau^3$ direction in color space) and utilizing the asymptotic
formulae listed at the end of Sec.~6 of \dkmone, one then obtains
the long-distance expressions
\def\trtwo{\tr^{}_2\,}
\def\ybar{\bar y}
\def\xbar{\bar x}
\eqna\ldholo 
$$\eqalignno{v_{mn}^{\sst\rm SD,LD}(y_i)\ &=\
{4i\over(y_i-x_0)^6}\,\sum_{k=1}^n\trtwo\wbar_k\tau^3w_k\,
(\ybar_i-\xbar_0)\,\sigma_{mn}\,(y_i-x_0)\ +\cdots
&{}\cr
&=\ 2i\pi^2\,G_{mn,pq}(y_i,x_0)\,\sum_{k=1}^n\trtwo\wbar_k\tau^3 w_k\,
\sigmabar^{pq}\ +\cdots\ ,
&\ldholo a
\cr
\lambda^\ld_\alpha(y_i)\ &=\ 4i\pi^2\,S_{\alpha\dalpha}(y_i,x_0)\,
\sum_{k=1}^n\wbar_k^{\dalpha\beta}\,(\tau^3)_\beta{}^\gamma\,\mu_{k\gamma}
\ +\cdots
\ ,&\ldholo b
\cr
\psi^\ld_\alpha(y_i)\ &=\ 4i\pi^2\,S_{\alpha\dalpha}(y_i,x_0)\,
\sum_{k=1}^n\wbar_k^{\dalpha\beta}\,(\tau^3)_\beta{}^\gamma\,\nu_{k\gamma}
\ +\cdots
\ ,&\ldholo c}
$$
omitting terms with a faster falloff.
Here $w_k,$ $\mu_k$ and $\nu_k$ are the top-row elements of the 
collective coordinate matrices
$a,$ $\M$ and $\N,$ respectively (see Eq.~\bcanonical);  the second
equality in Eq.~\ldholo a follows from Eq.~\Gpropertya.

We can now calculate, for example, the $n$-instanton contribution to the
effective \hbox{8-fermi} vertex
\eqn\eightfermi{{1\over 16} \ \psi^2\,\lambda^2\,\psibar^2\,\lambdabar^2\,
{\partial^4\over\partial\v^4}\,{\partial^4\over\partial\vbar^4}\,
\H(\v,\vbar)\ .}
Inserting 
\eqn\eightinsert{\psibar^\ld_\dalpha(x_1)\,\psibar^\ld_\dbeta(x_2)\,
\lambdabar^\ld_\dgamma(x_3)\,\lambdabar^\ld_\ddelta(x_4)\,
\psi^\ld_\alpha(y_1)\,\psi^\ld_\beta(y_2)\,
\lambda^\ld_\gamma(y_3)\,\lambda^\ld_\delta(y_4)}
into the collective coordinate integration and performing the $\xi^i$
integrals leaves
\def\dkappa{{\dot\kappa}}
\def\dlambda{{\dot\lambda}}
\def\drho{{\dot\rho}}
\def\dsigma{{\dot\sigma}}
\eqn\leftover{\eqalign{
&\int d^4x_0\, 
\epsilon^{\kappa\lambda} S_{\kappa\dalpha}(x_1,x_0)
S_{\lambda\dbeta}(x_2,x_0)\,
\epsilon^{\rho\sigma} S_{\rho\dgamma}(x_3,x_0)
S_{\sigma\ddelta}(x_4,x_0)
\cr
&\qquad\ \times\ 
\epsilon^{\dkappa\dlambda}S_{\alpha\dkappa}(y_1,x_0)S_{\beta\dlambda}(y_2,x_0)
\,\epsilon^{\drho\dsigma}S_{\gamma\drho}(y_3,x_0)S_{\delta\dsigma}(y_4,x_0)
\cr
&\times\ 
{\partial^4\over\partial\v^4}\,\int\Dmu\,e^{-\Sinstn}\,
\sum_{k,k',l,l'=1}^n\fourth(4i\pi^2)^4\,
(\nu^{}_k\tau^3w^{}_k\wbar^{}_{k'}\tau^3\nu^{}_{k'}) 
(\mu^{}_l\tau^3w^{}_l\wbar^{}_{l'}\tau^3\mu^{}_{l'})}
}
Again, we recognize this expression as the position-space Feynman graph
for a local $\psi^2\lambda^2\psibar^2\lambdabar^2$ vertex with an
effective coupling  given by the last line of Eq.~\leftover.
A comparison with the canonical form \eightfermi\ gives a formal expression
for the $n$-instanton contribution to $\Lfourderiv$, valid to
leading semiclassical order:
\eqn\Hndef{{\partial^4\over\partial\vbar^4}\,\H(\v,\vbar)\,
{\Big|}_{n\hbox{-}\rm inst}\ =\ 64 \pi^8
\int\Dmu\,e^{-\Sinstn}\,
\sum_{k,k',l,l'=1}^n\,
(\nu^{}_k\tau^3w^{}_k\wbar^{}_{k'}\tau^3\nu^{}_{k'}) 
(\mu^{}_l\tau^3w^{}_l\wbar^{}_{l'}\tau^3\mu^{}_{l'})}
This is the analog of Eq.~\Fndef\ for the prepotential.
Somewhat different (although necessarily consistent) formal expressions
for $\partial^2\H/\partial\vbar^2$ and $\partial^3\H/\partial\vbar^3$
may be derived in the same way, by examining the Green's functions
$\bG^{k,1}$ and $\bG^{k,2}$, respectively.
 Exchanging $\v$
and $\vbar$ in Eq.~\Hndef\ gives the $n$-antiinstanton contribution.
 There may also in general
be mixed \hbox{$n$-instanton}, $m$-antiinstanton contributions to $\cal H$
(unlike $\F$ due to holomorphicity), 
but these lie beyond the scope of this paper.

As a simple illustration, let us calculate the 1-instanton contribution
to $\H$ in the case of pure $N=2$ SYM theory. In that case the instanton
action reads \dkmone:
\eqn\simplest{S_{\rm inst}^{n=1}\ =\ 4\pi^2|\v|^2|w|^2-2\sqrtwo\,i\pi^2\vbar
\mu^\alpha\,(\tau^3)_\alpha{}^\beta\,\nu_\beta\ ,}
where $|\v|=\sqrt{\v\vbar}$. Also \dkmone:
\eqn\simplemeasure{\int\Dmu\ =\ {\Lambda^4 \over 2 \pi^4}
\int d^4w\,d^2\mu\,d^2\nu\ ,}
with $\Lambda$ the dynamically generated Pauli-Villars scale. The 
resulting integration in \Hndef\ is elementary, and gives\foot{Note
that only mixed derivatives of $\cal H$ with respect to both $\v$ and
$\vbar$ enter $\Lfourderiv$ so that $\cal H$ itself can be written in a
variety of equivalent ways.}
\eqn\simpleanswer{\H(\v,\vbar)\,{\Big|}_{1\hbox{-}\rm inst}\ = \
-{1 \over 8 \pi^2}{\Lambda ^4 \over \v^4}\, \log\vbar\ }
in accord with an earlier prediction of Yung's, arrived at using completely
different reasoning \Yung. In contrast, for $N_F>0,$ 
the first nonvanishing contribution to $\H$ is at the
2-instanton level, due to a discrete $\bigZ_2$ symmetry that forbids
all odd-instanton contributions \refs{\SW,\dkmfour}; it may be calculated
straightforwardly using the methods of \refs{\dkmone,\dkmfour}.

\bf 4. Nonrenormalization theorem for the $N=2,$ $N_F=4$
model. \rm To make further progress, we note a second general property of
$\Sinstn$ \refs{\dkmone,\dkmfour,\dkmsix}: 
when the hypermultiplet masses are zero, all dependence
on $\v$ and $\vbar$ can be eliminated from $\Sinstn$
by performing the collective coordinate rescaling
\eqn\rescaled{\eqalign{a&\rightarrow a/|\v|\ ,\quad
\M\rightarrow\M/\sqrt{\vbar}\ ,\quad
\N\rightarrow\N/\sqrt{\vbar}\ ,\quad\cr
\K&\rightarrow\K/\sqrt{\v}\ ,\quad
\Ktilde\rightarrow\Ktilde/\sqrt{\v}\ ,\quad
\R\rightarrow\R/\sqrt{\v}\ ,\quad
\Rtilde\rightarrow\Rtilde/\sqrt{\v}\ .}}
 Let us concentrate,
first, on the $N=2$ models with $0\le N_F\le4$ flavors of massless
fundamental hypermultiplets and no adjoint hypermultiplets. In these cases
the rescaling \rescaled\ implies 
\eqn\Dmuscale{\Dmu\ \rightarrow \ |\v|^{4-8n}(\sqrt{\vbar}\,)^{8n-4}
(\sqrt{\v}\,)^{2nN_F}\cdot\Dmu\ =\ 
\v^{2-(4-N_F)n}\cdot\Dmu}
so that
\eqn\Hscale{\H(\v,\vbar)\,{\Big|}_{n\hbox{-}\rm inst}
\ \sim\ {\log\vbar\over \v^{(4-N_F)n}}\ ,\qquad
\F(\v)\,{\Big|}_{n\hbox{-}\rm inst}
\ \sim\ {1\over \v^{(4-N_F)n-2}}\ }
as follows from Eqs.~\Hndef\ and \Fndef,  respectively.
In particular, for the special case $N_F=4,$ one has simply 
$\H\,{\big|}_{n\hbox{-}\rm inst}\,\sim\,\log\vbar$
so that the  effective component vertices contained in
$\Lfourderiv$ (all of which involve differentiating
 $\H$ with respect to both $\v$ and $\vbar$) automatically
vanish; likewise for the antiinstanton case with $\v\leftrightarrow\vbar.$
Thus we confirm the nonperturbative nonrenormalization theorem of
Dine and Seiberg in this model.\foot{Notice that, in 
contrast to $\H$,  in the $N_F=4$ model one also has
$\F\,{\big|}_{n\hbox{-}\rm inst}\,\sim\,\v^2$ so that the effective
$U(1)$ complexified coupling $\tau_{\rm eff}=\F''(\v)$
actually receives contributions from all (even) instanton 
numbers; see Refs.~\refs{\dkmfour,\dkmfive} for a discussion of this point.}
 Giving any of the hypermultiplets
a mass spoils the argument, since $m_i$ rescales to $m_i/\v$, and
this rescaled mass can be pulled down from the exponent.

\bf 5. Nonrenormalization theorem for the $N=4$ model. \rm
Next we consider the $N=4$ theory, i.e., $N=2$ SYM coupled to a single
massless adjoint hypermultiplet. In this model, after spontaneous
symmetry breakdown, the low-energy dynamics involves a larger set
of massless fields, corresponding to a single $N=4$ $U(1)$ multiplet.
Concomitantly, $\Sinstn$ is now independent of four additional
Grassmann collective coordinates: the `trace' components of the
$n\times n$ matrices $\R_\gamma'$ and $\Rtilde_\gamma'$ introduced in
Eq.~\newcanon\ \refs{\Seiberg,\dkmsix}. 
Respectively, these components constitute the third and
fourth supersymmetry modes, $\xi^3_\gamma$ and $\xi^4_\gamma.$ 
Now the collective coordinate integration takes the form
\eqn\foursusy{\int d^4x_0\,d^2\xi^1d^2\xi^2d^2\xi^3d^2\xi^4\,\B_n(\v,\vbar)\ ,}
where $\B_n$ is the $n$-instanton contribution to what one might
call the ``anteprepotential'' in analogy to Eq.~\Fndef:
\eqn\Bndef{\B_n(\v,\vbar)\ =\ \int\Dppmu\,e^{-\Sinstn}\ .}
Here $\Dppmu$ is the properly normalized integration measure for all collective
coordinates in the problem excepting the $N=4$ superspace position variables
$(x_0,\xi^1_\gamma,\xi_\gamma^2,\xi_\gamma^3,\xi_\gamma^4)$. As before, 
these eight unbroken  $\xi^i_\gamma$ modes
must be saturated by the insertion of an appropriate set
of fields, for instance the  eight antifermions
\def\chibar{\bar\chi}
\def\chitilde{\tilde\chi}
\def\chitildebar{\bar{\tilde\chi}}
\eqn\neweight{\bG^8(x_1,\ldots,x_8)
\ =\ \langle\psibar_\dalpha(x_1)\,\psibar_\dbeta(x_2)
\,\lambdabar_\dgamma(x_3)\,\lambdabar_\ddelta(x_4)\,
\chibar_\dkappa(x_5)\,\chibar_\dlambda(x_6)\,
\chitildebar_\drho(x_7)\,\chitildebar_\dsigma(x_8)\rangle\ ,}
where $\chi$ and $\chitilde$ are the adjoint hypermultiplet Higgsinos
associated with the collective coordinate matrices $\R$ and $\Rtilde,$
respectively. Now the action $\Sinstn$ in the $N=4$ model has the
discrete symmetry  $\{\M,\N,\v\}\leftrightarrow\{\R,\Rtilde,\vbar\}$
\dkmsix.
This symmetry, together with the long-distance expressions \LDfermions{},
implies
\eqna\LDchis
$$\eqalignno{
\chitildebar^\LD_\dalpha(x_i)
\ &=\ i\sqrtwo\,\xi^{3\alpha}\,S_{\alpha\dalpha}
(x_i,x_0)\,{\partial\over\partial\vbar}\ +\ \cdots
&\LDchis a
\cr
\chibar_\dalpha^\LD(x_i)
\ &=\ -i\sqrtwo\,\xi^{4\alpha}\,S_{\alpha\dalpha}
(x_i,x_0)\,{\partial\over\partial\vbar}\ +\ \cdots
&\LDchis b
}$$
{}From Eqs.~\LDfermions{} and \foursusy-\LDchis{} it follows that
 $\bG^8\big|_{n\hbox{-}\rm inst}\,\propto\,
\partial^8\B_n/\partial\v^4\partial\vbar^4$.
However, the rescaling argument \rescaled\ implies that 
\eqn\Dppmurescale{\Dppmu\ \rightarrow\
|\v|^{4-8n}(\sqrt{\vbar}\,)^{8n-4}(\sqrt{\v}\,
)^{8n-4}\cdot\Dppmu\ =\ \Dppmu\ ,}
 so that actually $\B_n(\v,\vbar)$ is a constant, 
independent of $\v$ and $\vbar$. Thus all (multi-)instanton contributions
to $\bG^8$ vanish, as predicted by Dine and Seiberg in this model as well.

Interestingly,
in the $N=8$ theory in three space-time dimensions,
obtained by dimensional reduction of the $N=4$ theory in 4D, this 
nonrenormalization theorem for the higher derivative terms no longer holds;
indeed the non-vanishing one-instanton 
contribution to the corresponding 4-derivative/8-fermion
term has been calculated in \PP, and all higher multi-instanton contributions
have been obtained in closed form in \dkmeight.
The significance of such instanton corrections for  Matrix theory
 was discussed last week in \BFSS.

\bigskip
\bigskip

ND is supported by a PPARC Advanced Research Fellowship, 
MPM  by the Department of Energy, and MJS and WAW 
by PPARC studentships.

\listrefs
\bye